\documentstyle[preprint,eqsecnum,aps]{revtex}
\begin{document}
\draft
\preprint{WU-JIK-96-2}
\title{Electrostriction of Polar Glasses}
\author{J. I. Katz}
\address{Department of Physics and McDonnell Center for the Space
Sciences\\Washington U., St. Louis, Mo. 63130\\katz@wuphys.wustl.edu}
\author{David R. Nelson}
\address{Lyman Laboratory of Physics\\Harvard U., Cambridge, Mass. 02138}
\date{\today}
\maketitle
\begin{abstract}
We develop a finite temperature theory for the susceptibility and
electrostriction of isotropic substances in which permanent electric dipoles
are restrained from free rotation by elastic forces.  All parameters are
constrained by the measured susceptibility and elastic constants.  When
applied to polyurethane, the predicted electrostriction is approximately
consistent with some of the wide range of data.  The saturation of the
electrostriction at high field may be explained qualitatively if the dipoles
consist of several amide groups locked together by crystallization of the
hard segments of the polymer.
\end{abstract}
\pacs{Pacs numbers: 77.22.Ch, 77.65.Bn, 77.84.Jd}
\narrowtext
\section{Introduction}
Zhenyi, {\it et al.} \cite{zhenyi} discovered that the electrostriction of
polyurethane film is large enough to be interesting.  The measurements of
Wang, {\it et al.} \cite{wang} found a smaller electrostriction.  The
experiments are unexpectedly difficult, partly because the material is
hygroscopic and its properties history-dependent, and partly because the
measured electrostriction is affected by the confinement of the soft
polyurethane by any stiffer material (such as electrodes) with which it is
in contact.

Comparatively little work exists \cite{anderson,sunder} on the theory of
electrostriction, and none directly applicable to this problem.  We develop
a simple theory of the electric susceptibility of polymers (such as
polyurethane) with polar groups, and calculate electrostriction from the 
susceptibility.  The theory has one parameter, which may be determined by
comparing to the observed susceptibility.
\section{Susceptibility}
Polyurethane elastomers are block copolymers containing crystallizable and
noncrystallizable segments \cite{zhenyi}.  The molecular structure of a
generic polyurethane is shown in Figure \ref{fig1}, with $x$ typically
about 2.  The crystallizable (hard) segments have attached amide groups with
permanent electric dipole moments $p \approx 3.6 \times 10^{-18}$ esu-cm.
Our model for the susceptibility considers the orientation of the dipoles in
the electric field, and for the moment neglects the intrinsic susceptibility
of the nonpolar portions of the polyurethane.  For comparison, nonpolar
polymers such as polyethylene typically have dielectric constants (resulting
entirely from polarization of the chemical bonds) of about 2.3, while
polyurethanes have dielectric constants of about 6.8.

We make an independent particle approximation in which the dipoles in
electrified polyurethane have two contributions to their energy---the usual
electrostatic energy 
\begin{equation}
U_1 = -{\bf p}\cdot{\bf E}, \label{eq1}
\end{equation}
and a term resulting from their rotation from their equilibrium
orientations.  This latter energy results from the elastic restoring forces
of the polymer chains and of the bulk medium (entangled chains).  For small
deviations $\Delta \theta \ll 1$ from the equilibrium position the elastic
energy will, in general, be quadratic in $\Delta \theta$.  We therefore write
\begin{equation}
U_2 = {1 \over 2} S  (\Delta \theta)^2, \label{eq2}
\end{equation}
where $S$ is an elastic stiffness constant.  For simplicity we assume that
$S$ is a scalar so that $U_2$ depends only on the magnitude of the angular
deflection, and not on its direction; this may not be true for dipoles on
a polymer, but in the absence of a detailed model for the orientation of the
amide groups in polyurethane it is a reasonable approximation.  If the
applied electric field defines the $\theta=0$ axis of a polar coordinate
system, and the orientation of the dipole defines the $\phi=0$ plane and
makes an angle $\theta$ to the electric field, then $\Delta \theta$ is
obtained from
\begin{equation}
\cos \Delta \theta = \sin \theta_0 \sin \theta \cos \phi_0 + \cos \theta_0
\cos \theta, \label{eq3}
\end{equation}
where $(\theta_0, \phi_0)$ is the equilibrium position of the dipole in the
absence of a field.

Unfortunately \ref{eq2} becomes cumbersome when \ref{eq3} is substituted,
and the integrals required to take thermal averages cannot be done
analytically.  Therefore, we replace \ref{eq2} by
\begin{equation}
U_2 = S(1-\cos \Delta \theta). \label{eq4}
\end{equation}
To second order in $\Delta \theta$ \ref{eq4} agrees with \ref{eq2}.  Because
the actual calculation of $U_2$ for large $\Delta \theta$ is a problem in
nonlinear elasticity, which would be quite difficult, there is no reason to
think that \ref{eq4} is any less accurate than \ref{eq2}.  The form
\ref{eq4} may be rewritten
\begin{equation}
U_2 = S  - {\bf  p}\cdot{\bf \tilde E}, \label{eq5}
\end{equation}
where $\bf \tilde E$ is a vector with the dimensions of an electric field
which points in the direction of the equilibrium position of the dipole (for
${\bf E} = 0$), and which has magnitude ${\tilde E} = S/p$.  The effects of
the elastic forces on the dipole have now been incorporated into the field
$\bf \tilde E$, which will be assumed to be an isotropic random function of
space with zero correlation length and fixed modulus $\tilde E$.  The total
energy $U = U_1 +U_2$ may be obtained from \ref{eq1} and \ref{eq5}, dropping
the constant term in \ref{eq5}:
\begin{equation}
U = -{\bf p}\cdot({\bf E} + {\bf \tilde E}). \label{eq6}
\end{equation}

We ignore the electrostatic and elastic interaction between dipoles, in
effect assuming their density to be low.  The physics of randomly oriented
dipoles embedded in an amorphous elastic polymer matrix is described by the
random field $\bf \tilde E$.  If interactions between dipoles were to be
included the problem would be that of a dipole glass, analogous to the
familiar (but difficult) spin glass problem.

We evaluate the mean polarization $\langle {\bf p} \rangle$ in thermal
equilibrium at temperature $T$.  For a given dipole the energy \ref{eq6}
leads, by the usual elementary calculation, to a polarization of magnitude 
\begin{equation}
\vert \langle {\bf p} \rangle_T \vert = L(x_c) p 
\end{equation}
along the direction of ${\bf E} + {\bf \tilde E}$, where the Langevin
function
\begin{equation}
L(x) \equiv \coth(x) - {1 \over x},
\end{equation}
the combined normalized effective field strength is
\begin{equation}
x_c \equiv {\vert {\bf E} + {\bf \tilde E} \vert \over kT},
\end{equation}
and $\langle \rangle_T$ denotes a thermal average only.

The net polarization of the medium must, by symmetry, be in the direction
($\bf \hat z$) of the applied field $\bf E$, and is found from
\begin{equation}
{\overline{\langle p_z \rangle_T} \over p} = {1 \over 4 \pi} \int L(x_c)
\cos \theta^\prime d\Omega,
\end{equation}
where the overline represents a quenched angular average over all directions
of $\bf \tilde E$ and $\theta^\prime$ is the angle between $\bf E$ and ${\bf
E} + {\bf \tilde E}$.  Upon expanding $L(x_c)$ and $\cos \theta^\prime$ in
powers of $\delta \equiv E/{\tilde E}$ and $\xi \equiv pE/kT = x_0\delta$,
where $x_0 \equiv p {\tilde E}/kT$, and keeping only terms of first order in
$E$, we find
\begin{equation}
{\overline{\langle p_z \rangle_T} \over p} = {2 \over 3} \delta L(x_0) + {1
\over 3} \xi L^\prime(x_0). \label{eq11}
\end{equation}
Note that while $\delta \ll 1$ and $\xi \ll 1$ are assumed, $x_0$ is
generally not small.  For a density $n$ of independent dipoles the
macroscopic polarization ${\bf P} = n\overline{\langle{\bf p}\rangle_T}$ may
be obtained from \ref{eq11}:
\begin{equation}
{\bf P} = np{{\bf E} \over {\tilde E}} \left[{2 \over 3} L(x_0) +
{1 \over 3} x_0 L^\prime (x_0) \right] \equiv np {{\bf E} \over {\tilde
E}} f(x_0). \label{eq12}
\end{equation}
The function $f(x)$ is plotted in Figure \ref{fig2}.  The limits of
\ref{eq12} are
\begin{equation}
{\bf P} \approx \cases{{\displaystyle np^2{\bf E} \over \displaystyle 3kT},
&if $x_0 \ll 1$;\cr {\displaystyle 2np{\bf E}\over \displaystyle
3{\tilde E}},& if $x_0 \gg 1$.\cr}
\end{equation}
The limit for $x_0 \ll 1$ is a familiar elementary result, while the limit
for $x_0 \gg 1$ was given in slightly different form by Fr\"olich
\cite{frolich}.

Polyurethanes typically have dielectric constants $\epsilon \approx 6.8$.
Using the relation (in cgs units) $\epsilon = 1 +4 \pi \chi$, this yields a
susceptibility $\chi = 0.46$.  The self-consistent molecular polarizability
$\gamma$ (here a scalar, but in general a tensor) is 
\begin{equation}
{p \over {\tilde E}} f\left({p{\tilde E} \over kT}\right) =
{\overline{\langle p_z \rangle_T} \over E} \equiv \gamma = {3 \over 4 \pi n}
\left( {\epsilon - 1 \over \epsilon + 2} \right), \label{eq14}
\end{equation}
where the last equality is the Clausius-Mossotti equation \cite{jackson}.
This equation may be modified straightforwardly to allow for a dielectric
constant $\epsilon_c$ of the continuum in which the dipoles are embedded.
The result is
\begin{equation}
{p \over {\tilde E}} f\left({p{\tilde E} \over kT}\right) = \gamma = {3
\over 4 \pi n} \left({\epsilon - \epsilon_c \over \epsilon + 2}\right).
\label{eq15}
\end{equation}
For the polyurethane shown in Figure \ref{fig1} with $x=2$ and density 1.1
gm/cm$^3$, the dipole density $n = 3.4 \times 10^{21}$ cm$^{-3}$, assuming
each amide group rotates independently.  Then, with $p = 3.6 \times
10^{-18}$ esu-cm, the self-consistent solution to the transcendental
equation \ref{eq15}, taking $T =  300^{\circ}$K and $\epsilon_c = 2.3$, is
$p{\tilde E}/kT = 5.2$, {\it i.e.} ${\tilde E} = 6.1 \times 10^4$ cgs ($=
1.8 \times 10^7$ V/cm).  It should be remembered that these numerical values
depend on the assumed numerical values of $n$ and $p$, which are poorly
known.

The twisting stiffness constant equivalent to the derived value of
$\tilde E$ is $S = {\tilde E} p = 2.2 \times 10^{-13}$ erg.  This may be
compared to the macroscopic elastic constants of polyurethane, which
typically has a Young's modulus $Y \approx 3 \times 10^8$ dyne/cm$^2$.  If
the dipole is described by a rigid sphere of radius $a$, then rotation by an
angle $\Delta\theta$ implies a strain $\sim\Delta\theta$, a tangential
stress on the surface of the sphere $\sim Y \Delta \theta$, and a total
torque $\sim 2 \pi Y a^3 \Delta \theta$.  The corresponding stiffness
constant $S \approx 2 \pi Y a^3$, so that $a \approx 5$\AA.  This value is
not unreasonable, although perhaps a bit large, but its significance is
uncertain: The dipoles are mechanically coupled to their environment in
complex ways.  They are parts of a nearly inextensible and flexurally stiff
covalently bonded polymer chain, which can shear quite easily with respect
to its neighbors, except as limited by entanglements.  These elastic
properties are strongly temperature dependent, so that $\tilde E$ will also
be sensitive to temperature, at least for temperatures of order room
temperature.  Most of the temperature dependence of \ref{eq11} and
\ref{eq12} is likely to come from the variation of $\tilde E$ with $T$,
rather than from the explicit dependence on $T$, except perhaps at very low
temperatures.  In fact, some polar polymers are known \cite{aip} to have
values of $\epsilon$ which increase rapidly with $T$, as this argument would
suggest.
\section{Saturation}
The derivation of \ref{eq12} is limited to small electric fields.  The
effective electric field $E_{eff}$ acting on an object in an amorphous
medium is (in cgs units)
\begin{equation}
E_{eff} = E  +  {4 \pi \over 3} P = E \left({\epsilon + 2 \over 3} \right)
\approx 2.9 E,
\end{equation}
where we have taken the empirical dielectric constant of polyurethane.  In
order for \ref{eq12} to be valid the conditions $pE_{eff} \ll kT$ and
$E_{eff} \ll {\tilde E}$ must both be met; when they are not met $\langle
p_z \rangle$ becomes a significant fraction of $p$, the polarization
saturates, and $\epsilon$, $\chi$, and the electrostriction decrease below
the values given by the low field theory.  The available strong-field
data\cite{zhenyi} only describe the electrostriction, and show that it
saturates at $E \approx 10^7 {\rm V/m} \approx 300$ cgs,  corresponding to
$E_{eff} \approx 900$ cgs, $pE_{eff}/kT \approx 0.1$, and
$E_{eff}/{\tilde E} \approx 0.02$.  We expect the susceptibility to
saturate at the same field strengths as does the electrostriction, a
prediction which is readily tested.

The electrostriction saturates at fields substantially smaller than
expected.  This discrepancy may be resolved if the actual rotating dipole
moments consist of a number of individual amide groups, locked together by
crystallization of the hard segments of the polymer.  Such a rigid block
would have a value of $p$ several times that of an individual amide group,
and $n$ would represent the density of such blocks and would be
correspondingly smaller than the density of individual amide groups.  The
value of $np$ would not be greatly changed by this locking of individual
amides into a single giant dipole.  Because $f(x)$ is a slowly varying
function for $x \gg 1$, $\tilde E$ would also not be greatly changed.
\section{Electrostriction}
The Landau theory of electrostriction defines a free energy per unit volume
of a body in an imposed stress field $\sigma_{ij}$ and electric field $E_i$:
\begin{equation}
F(\sigma_{ij},E_i) = \mu u_{ij}^2 + {1 \over 2} \lambda u_{mm}^2  -
u_{ij}\sigma_{ij} + {1 \over 2} (\chi^{-1})_{ij}P_i P_j + Q_{ijkl}P_i P_j
u_{kl} - E_i P_i, \label{eq19}
\end{equation}
where $\mu$ and $\lambda$ are Lam\'e coefficients of elasticity, $u_{ij}$ is
the strain tensor, $\sigma_{ij}$ is the stress tensor, $(\chi^{-1})_{ij}$ is
the inverse of the susceptibility tensor $\chi_{ij}$ defined by $P_i =
\chi_{ij} E_j$, $Q_{ijkl}$ is the electrostriction tensor, and $P_i$ and
$E_i$ are the polarization and electric field.  In this theory $u_{ij}$ and
$P_i$ describe the internal state of the material, and characterize its
response to the externally imposed parameters $\sigma_{ij}$ and $E_i$.  If
we neglect fluctuations, which are small far from phase transitions,
$u_{ij}$ and $P_i$ may be found by minimizing \ref{eq19}.  

By multiply differentiating the Legendre transform of \ref{eq19}
\begin{equation}
G(\sigma_{ij},E_i) \equiv F + u_{ij}\sigma_{ij} + E_i P_i
\end{equation}
in various orders we can relate $Q_{ijkl}$ to the derivatives of
$(\chi^{-1})_{ij}$ with respect to $u_{kl}$:
\begin{equation}
Q_{ijkl} = {\partial \over \partial P_i}{\partial \over \partial P_j}
{\partial G \over \partial u_{kl}} = {\partial \over \partial u_{kl}}
{\partial \over \partial P_i}{\partial G \over \partial P_j},
\end{equation}
{\it i.e.,}
\begin{equation}
Q_{ijkl} = {\partial \over \partial P_i}{\partial \over \partial P_j}
\sigma_{kl} = {\partial \over \partial u_{kl}} (\chi^{-1})_{ij},\label{eq21b}
\end{equation}
where \ref{eq21b} has been obtained by using the thermodynamic relations
$\partial G / \partial u_{kl} = \sigma_{kl}$ and $(\chi^{-1})_{ij} =
\partial^2 G / \partial P_i \partial P_j$.  We then use our elementary
microscopic model of $\chi$ to calculate $Q_{ijkl}$ from the second part of
\ref{eq21b}: $Q_{ijkl} = \partial (\chi^{-1})_{ij} / \partial u_{kl}$.

The application of a strain component $u_{33}$ to an isotropic distribution
of vectors $\bf \tilde E$ (the equilibrium orientations of the dipoles)
rotates them.  We assume that the magnitude $\tilde E$ is not changed by the
application of the strain field.  We do not know if this is true, but it
could not be described except by a completely phenomenological parameter,
which we have no independent means of determining.  The angle between $\bf
\tilde E$ and the $\theta = 0$ axis is rotated by an amount, found from
elementary geometry
\begin{equation}
\Delta \theta = -u_{33} \sin \theta \cos \theta.
\end{equation}
The calculation of susceptibility is then repeated, replacing $\theta$ by
$\theta + \Delta \theta$, with the result, correct to first order in
$u_{33}$,
\begin{equation}
\gamma_{33} \equiv {\overline{\langle p_z \rangle_T} \over E_z} = {p \over
{\tilde E}}[f(x_0) + g(x_0)u_{33}],
\end{equation}
where the function $g(x)$ is defined
\begin{equation}
g(x) \equiv -{4 \over 15}[L(x) - xL^{\prime}(x)],
\end{equation}
and is plotted in Figure \ref{fig2}.  By considerations of symmetry (or
explicit calculation) it is seen that $\gamma_{ij} = \partial
\overline{\langle p_i \rangle_T} / \partial E_j$ is diagonal, as is
$\chi_{ij}$, so that $(\chi^{-1})_{33} = (\chi_{33})^{-1}$.  Then, to lowest
order in $u_{33}$, noting that $\Delta n = -u_{33}n$ and $\partial
(\chi^{-1})_{ij} / \partial (n \gamma^{-1})_{ij} = 1$, we have
\begin{equation}
Q_{3333} = {\partial (\chi^{-1})_{33} \over \partial u_{33}} \approx
{{\tilde E} \over npf(x_0)}\left(1 - {g(x_0) \over f(x_0)}\right).
\label{eq25}
\end{equation}
We have neglected the dependence on strain of the susceptibility of the
continuum medium in which the dipoles are embedded; any such dependence is
likely to be very small, especially in a soft elastomer in which
deformations are almost purely volume-conserving.

The effects of strain components $u_{11}$ and $u_{22}$ may be found
similarly:
\begin{equation}
\Delta \theta = (u_{11}\sin^2\phi + u_{22}\cos^2\phi)\sin\theta\cos\theta,
\end{equation}
where $\phi$ is the azimuthal angle of $\bf \tilde E$, measured from the
y-axis.  Then
\begin{equation}
Q_{3311} = Q_{3322} = {{\tilde E} \over npf(x_0)}\left(1 + {g(x_0) \over 2
f(x_0)}\right). \label{eq27}
\end{equation}
Similar calculations for $u_{12}$, $u_{13}$, and $u_{23}$ (or symmetry
arguments) show that
\begin{equation}
Q_{33kl} = 0 \quad {\rm for}\ k \ne l. \label{eq28}
\end{equation}
Similarly, because the distribution of $\bf \tilde E$ retains inversion
symmetry in the strained state in all of these strained configurations
$\chi_{ij}$ remains diagonal, so that $Q_{ijkl} = 0$ for $i \ne j$.
Permutation of indices leads to the general result
\begin{equation}
Q_{ijkl} = \cases{\displaystyle{{\tilde E} \over npf(x_0)}\left(1 - {g(x_0)
\over f(x_0)}\right) & for $i=j=k=l$;\cr
\displaystyle {{\tilde E} \over npf(x_0)}\left(1 + {g(x_0) \over
2f(x_0)}\right) & for $i=j\ne k=l$;\cr 0 & otherwise.\cr}
\end{equation}

We now calculate the electrostrictive strain produced by the application of
an electric field and polarization in the $z$ direction.  Differentiating
\ref{eq19} with respect to $u_{kl}$, and setting the derivative equal to
zero in equilibrium yields
\begin{equation}
0 = 2 \mu u_{kl} + \delta_{kl} \lambda u_{mm} - \sigma_{kl} + Q_{ijkl}P_i
P_j. \label{eq30}
\end{equation}
The electrostrictive term acts as an effective stress
\begin{eqnarray}
\sigma^{\prime}_{kl}&= -Q_{ijkl}P_i P_j \\ &= -Q_{33kl}P_3^2. \label{eq31a}
\end{eqnarray}
Note that linearizing \ref{eq31a} about a state of nonzero ${\bf E} = (E_0
+ \Delta E){\bf {\hat z}}$ leads to
\begin{equation}
\Delta \sigma^\prime_{kl} = - 2Q_{33kl}{\partial P(E_0) \over \partial E}
P(E_0) \Delta E;
\end{equation}
{\it i.e.}, an effective piezoelectric response.
Using \ref{eq25}, \ref{eq27} and \ref{eq28} yields
\begin{equation}
\sigma^{\prime}_{11} = \sigma^{\prime}_{22} = -{{\tilde E} \over
npf(x_0)}\left(1 + {g(x_0) \over 2f(x_0)}\right)P_3^2
\end{equation}
\begin{equation}
\sigma^{\prime}_{33} = -{{\tilde E} \over npf(x_0)}\left(1 - {g(x_0) \over
f(x_0)} \right)P_3^2;
\end{equation}
the off-diagonal components are zero.  Substitution in the elastic
stress-strain relation (or use of \ref{eq30}) yields the strain
\begin{equation}
u_{11} = u_{22} = \left(-{{\tilde E} \over 4 \mu np}{g(x_0) \over
f^2(x_0)} + {{\tilde E} \over 3Knp} {1 \over f(x_0)}\right)P_3^2
\end{equation}
\begin{equation}
u_{33} = \left({{\tilde E} \over 2 \mu np}{g(x_0) \over f^2(x_0)} +
{{\tilde E} \over 3Knp}{1 \over f(x_0)}\right)P_3^2, \label{eq33}
\end{equation}
where the bulk modulus $K = \lambda + {2 \over 3}\mu$.

In order to rewrite \ref{eq33} in terms of $E_3$ the Clausius-Mossotti
equation \ref{eq14}, \ref{eq15} may be solved for $\chi$ in terms of
$f(x_0)$, but because $\chi$ is measured it is easier simply to substitute
$P = \chi E$: 
\begin{equation}
u_{11} = u_{22} = \left(-{{\tilde E} \over 4 \mu np}{g(x_0) \over
f^2(x_0)} + {{\tilde E} \over 3Knp} {1 \over f(x_0)}\right)\chi^2 E_3^2
\end{equation}
\begin{equation}
u_{33} = \left({{\tilde E} \over 2 \mu np}{g(x_0) \over f^2(x_0)} +
{{\tilde E} \over 3Knp}{1 \over f(x_0)}\right)\chi^2 E_3^2. \label{eq34}
\end{equation}

For materials which are soft to shear, with $\mu \ll K$ (as in many
polymers) the second terms in parentheses of \ref{eq33} and \ref{eq34} are
small compared to the first terms.  It is evident that materials with small
shear moduli $\mu$ (or small Young's moduli, equal to $3 \mu$ in the limit
$\mu/K \rightarrow 0$) will show comparatively large electrostrictive
strains.  In addition, even though typically $Q \propto \chi^{-1}$, the
strain $u \propto \chi$ and materials with large $\chi$ (but small $Q$) are
usually the best electrostrictors \cite{newnham}.
\section{Comparison to Experiments}
Substitution in \ref{eq34} for PT6100S polyurethane \cite{zhenyi}, which has
$\mu \approx 3 \times 10^7$ erg/cm$^3$, taking $K \gg \mu$, and using the
estimated $n$ and $p$ and the value of $\tilde E$ previously determined from
the susceptibility, yields $u_{33} = - 7 \times 10^{-9}\ {\rm cm}^4/
{\rm esu}^2\ E_3^2$, where the esu is the cgs unit of charge ($1/(3 \times
10^9)$ Coulomb).  Zhenyi, {\it et al.} \cite{zhenyi} find $u_{33} \approx - 2
\times 10^{-8}\ {\rm cm}^4/ {\rm esu}^2\ E_3^2$, but with a range of values
of a factor $\sim 2$ depending on the details of sample preparation and of
mechanical constraint during the experiments.

Wang {\it et al.} \cite{wang} find for 80CF-2 polyurethane $u_{33} \approx
- 3 \times 10^{-9}\ {\rm cm}^4/{\rm esu}^2\ E_3^2$ (approximately 40\% of
the effect results from Maxwell stress rather than electrostriction).  For
this material $\mu \approx 1.0 \times 10^8$ erg/cm$^3$ \cite{wang} so that
we predict (\ref{eq34}) $u_{33} \approx - 2.4 \times 10^{-9}\ {\rm cm}^4/
{\rm esu}^2\ E_3^2$, approximately consistent with experiment.  The result
\ref{eq34} contains no free parameters because the ratio ${\tilde E}/(np)$
is determined from the measured dielectric constant by \ref{eq15}, as is
$\chi$, and $\mu$ is directly measured.
\acknowledgments
We thank E. Balizer, J. Barger, L. E. Cross, A. Ellinthorpe, V. Kugel, N.
Lewis, R. E. Newnham, J. I. Scheinbeim and H. Wang for discussions.  This
work was supported in part by DARPA at Washington University and by NSF DMR
94-17047 at Harvard University.
\appendix
\section{Volume Torques}
It is evident that there are no body forces on a uniformly electrified
homogeneous material, but it is still of interest to examine the strain
field $\bf u$ produced by a point torque in an elastic medium, such as the
permanent dipoles in electric fields discussed in this paper, in order to
see if this strain contributes to the electrostriction.  This result
not readily found in the elasticity literature.

Begin with Kelvin's result for the displacement field $\bf u$ produced by a
point force $\bf F$ at the origin of an elastic medium \cite{landau}:
\begin{equation}
{\bf u}({\bf r}) = \left({1 + \sigma \over 8 \pi Y (1 - \sigma)}\right)
\left({(3 - 4 \sigma){\bf F} + {\bf \hat n}({\bf \hat n}\cdot{\bf F}) \over
r}\right), \label{eqa1}
\end{equation}
where $\sigma$ is Poisson's ratio, $\bf r$ denotes the point at which the
displacement is measured, and $\bf \hat n$ is the unit vector in the
direction of $\bf r$.  By adding to \ref{eqa1} the displacement produced
by a force $-{\bf F}$ acting at a point $\bf \Delta r$ we find the
displacment produced by a point torque ${\bf N}  = - {\bf \Delta r} \times
{\bf F}$:
\begin{equation}
{\bf u}({\bf r}) = \left({1 + \sigma \over 8 \pi Y (1 - \sigma)}\right)
{{\bf N} \times {\bf r} \over r^3}.
\end{equation}

A dipole $\bf p$ in an electric field $\bf E$ is subject to a torque
\begin{equation}
{\bf N} = {\bf p} \times {\bf E}.
\end{equation}
In a medium with an inversion-symmetric distribution of dipole orientations
$\langle {\bf p} \rangle = 0$ and $\langle {\bf N} \rangle = 0$.  Each
dipole is rotated by an angle ${\bf \Delta \theta} = {\bf N}/S$.

The rotation of the dipoles changes the direction of their moments, and
therefore the torque.  The change in $\bf p$ is 
\begin{equation}
{\bf \Delta p} = {\bf \Delta \theta} \times {\bf p} = {{\bf N} \times
{\bf p} \over S} = {({\bf p} \times {\bf E}) \times {\bf p} \over S} =
{{\bf E}p^2 -  {\bf p}({\bf p}\cdot{\bf E}) \over S},
\end{equation}
and yields an incremental torque, second order in $\bf E$ (like the
electrostriction):
\begin{equation}
{\bf N} = {\bf \Delta p} \times {\bf E} = -{({\bf p}\times{\bf E})({\bf p}
\cdot{\bf E}) \over S}. \label{eqa5}
\end{equation}
Although \ref{eqa5} is second order in both $\bf p$ and $\bf E$, in any
volume element it will average to zero for an inversion-symmetric
distribution of $\bf p$, and hence it is not a useful model of
electrostriction.
\section{Correlation of Electrostriction with Thermal Expansion}
Uchino, {\it et al.} \cite{uchino} note that the electrostriction
coefficient $Q$ and the thermal expansion coefficient $\alpha$ are
empirically related by $Q \propto \alpha^2$ for a wide variety of materials,  
including one polymer (PVDF), pyrex glass, ceramic piezoelectrics, and ionic
crystals (see their Figure 1); there is only a factor of 2 (in $\alpha$) or
4 (in Q) scatter about the power law fit, which extends over a range of
about 30 in $\alpha$ (1000 in $Q$).  This empirical relation has a very
elementary explanation.

The simplest model of thermal expansion is a particle in a weakly anharmonic
potential $V = ax^2 + bx^3$.  If the particle has an energy $kT$, its
turning points are shifted by $\Delta x \approx - {b kT / (2 a^2)}$, giving a
linear thermal expansion coefficient $\alpha = {\vert b \vert / (2 a^2 a_0)}$,
where $a_0$ is the distance between the potential minimum and the origin (or
the nearest neighboring particle).

The analogous simple model of electrostriction considers two particles in
the well, with charges $\pm q$, and adds an electric field $E$.  The
particles are now in potentials $V_\pm = ax_\pm^2 + bx_\pm^3 \pm qEx_\pm$,
and their equilibrium ($T=0$) positions are given by $x_\pm = {-a +
\sqrt{a^2 \mp  3bqE} / (3b)}$.  The displacement of their barycenter is
$\Delta x = (x_+ + x_-)/2 \approx -{3bq^2 E^2 / (8a^4)}$, giving an
electrostrictive coefficient $Q = -{3bq^2 / (8a^4a_0)}$ (note that $b<0$ for
realistic potentials).

If the chief variation between materials is in $a$ then $Q \propto
\alpha^2$, as observed; the scatter about this relation arises from
variations in $b$, $a_0$, and the effective charge $q$.  The parameter $a$
is related to the bulk modulus, which varies a great deal between materials; it is not surprising that variations in the other variables are less important,
though not insignificant.
\section{Electrostriction in the Melt}
It is possible to construct a very simple model of electrostriction of a
melted poly\-urethane, with no elastic resistance to rotation of the
permanent dipoles.  Suppose the polymer configuration to be a random walk
which deviates from isotropy because an applied electric field distorts the
equilibrium distribution of the orientations of the dipoles (for small
anisotropy the self-avoiding property of the walk will not change the degree
of anisotropy).  Consider dipole components $p_\perp$ perpendicular to a
fraction $f$ of the units of the polymer chain, but otherwise free to
rotate.  The distribution of dipole orientations making an angle $\theta$
with respect to an electric field in the z direction is given by the
elementary result
\begin{equation}
P(\theta) = {v e^{-v\cos\theta} \over 2 \sinh v},
\end{equation}
where $v \equiv p_\perp E_{eff}/kT$ and $E_{eff} = (1 + {4\pi \over 3}\chi)E$
is the effective field at a dipole in a random medium.  Each polymer ${\bf
r}(s)$ can now be regarded as a biased random walk, whose mean square
end-to-end distance grows anisotropically as a function of arc length
$\langle[r_i(s) - r_i(0)][r_j(s) - r_j(0)]\rangle = (2 D_{ij} s)^{2\eta}$.
For chain elements of unit length the random walk is described by a diagonal
diffusion tensor, correct to second order in $u$,
\begin{equation}
D_{xx} = {1 \over 2}\left(\langle y^2_d \rangle + \langle z^2_d
\rangle\right) = {1 \over 3} \left(1 + {f u^2 \over 30}\right)
\end{equation}
\begin{equation}
D_{yy} = {1 \over 2}\left(\langle x^2_d \rangle + \langle z^2_d
\rangle\right) = {1 \over 3} \left(1 + {f u^2 \over 30}\right)
\end{equation}
\begin{equation}
D_{zz} = {1 \over 2}\left(\langle x^2_d \rangle + \langle y^2_d
\rangle\right) = {1 \over 3}\left(1 - {f u^2 \over 15}\right),
\end{equation}
where the subscript $d$ refers to the components of the dipole vectors.
The corresponding strain is
\begin{equation}
u_{zz} = - {f \eta\over 15}\left({p_{\perp} \over kT}\right)^2 \left(1 +
{4 \pi \over 3} \chi\right)^2 E^2 = - 2.7 \times 10^{-9} E^2
\end{equation}
\begin{equation}
u_{xx} = u_{yy} = - {u_{zz} \over 2},
\end{equation}
where we have taken $f = 0.2$, $\eta = 0.6$, $p_{\perp} = 2.5 \times
10^{-18}$ esu-cm, $T = 300^{\circ}$K, and $\chi = 0.46$ (corresponding to
{\it solid} polyurethane with $\epsilon=6.8$).  This strain will make the
material optically active, but in order to calculate this it would be
necessary to know the intrinsic dielectric anisotropy of the polymer chain
at optical frequencies; optical activity is determined by the strain alone,
and is not specific to its electrostrictive origin.

\begin{figure}
\caption{Molecular structure of polyurethane elastomer.}
\label{fig1}
\end{figure}
\begin{figure}
\caption{The functions $f(x)$ and $g(x)$}
\label{fig2}
\end{figure}
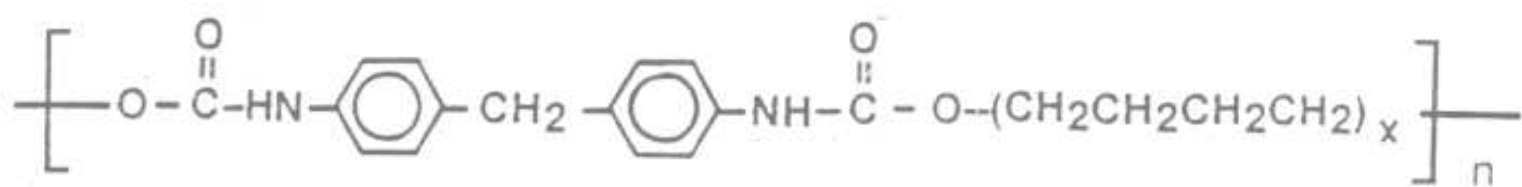
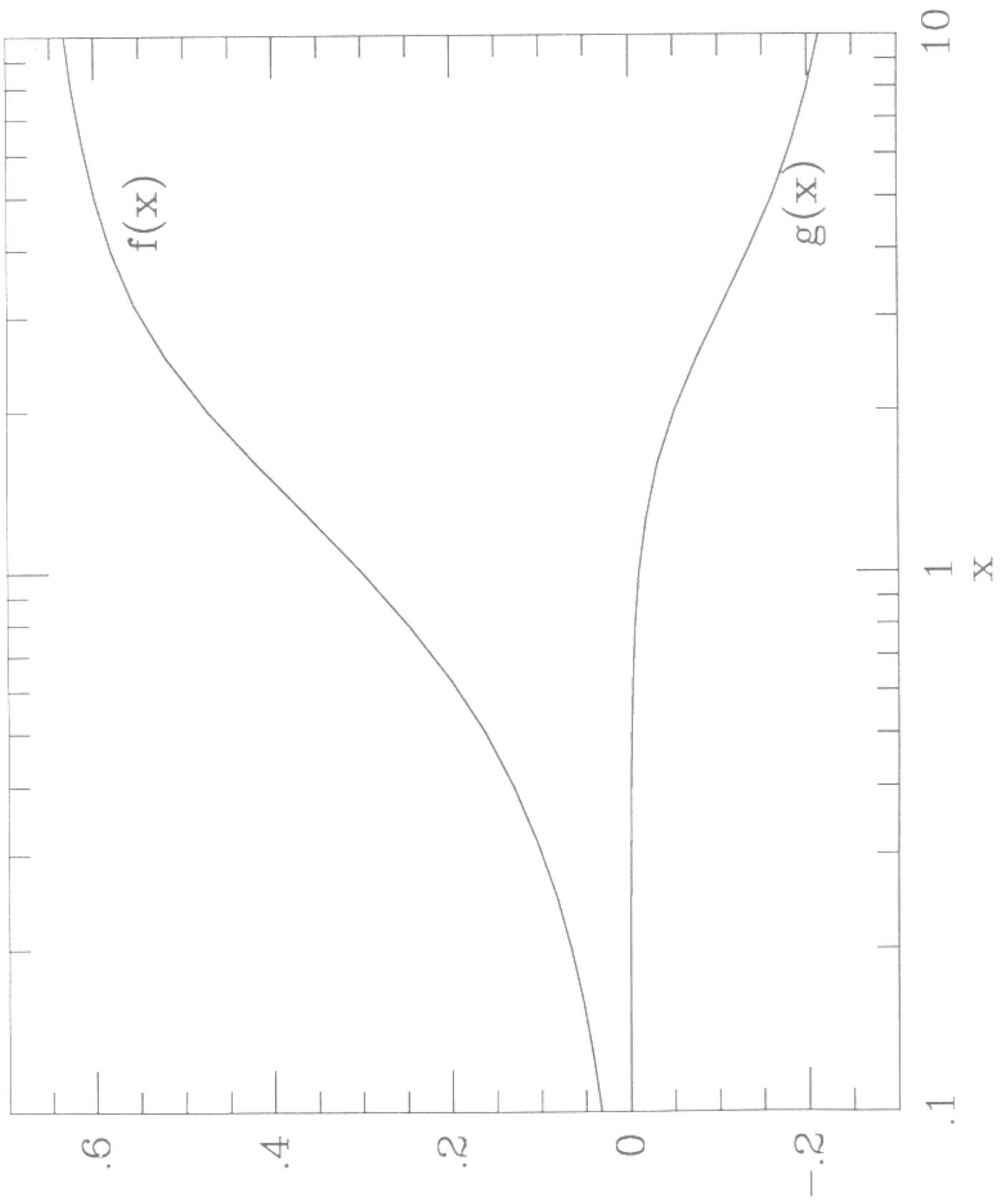
\end{document}